\documentclass[12pt,letterpaper]{article}
\pdfoutput=1

\usepackage{cite}
\usepackage{graphicx,array}
\usepackage{color}
\usepackage{latexsym}
\usepackage{amsthm}
\usepackage{amsmath}
\usepackage{amssymb}
\usepackage{hyperref}
\usepackage{bbold}
\usepackage{subfig}
\usepackage{multirow}

\usepackage{hyperref}

\def\simgt{\mathrel{\lower2.5pt\vbox{\lineskip=0pt\baselineskip=0pt
           \hbox{$>$}\hbox{$\sim$}}}}
\def\simlt{\mathrel{\lower2.5pt\vbox{\lineskip=0pt\baselineskip=0pt
           \hbox{$<$}\hbox{$\sim$}}}}

\newcommand{\be}{\begin{equation}}
\newcommand{\ee}{\end{equation}}
\newcommand{\bea}{\begin{eqnarray}}
\newcommand{\eea}{\end{eqnarray}}

\newcommand{\GeV}{\textrm{ GeV}}
\newcommand{\TeV}{\textrm{ TeV}}
\newcommand{\gsim}{\lower.7ex\hbox{$\;\stackrel{\textstyle>}{\sim}\;$}}
\newcommand{\lsim}{\lower.7ex\hbox{$\;\stackrel{\textstyle<}{\sim}\;$}}

\newcommand{\BR}{\mathcal{B}}

\definecolor{darkblue}{cmyk}{1,0.3,0,0.2}
\definecolor{violet}{cmyk}{0,1,0,0.2}
\hypersetup{colorlinks, bookmarksnumbered, citecolor=darkblue, linkcolor=darkblue, pdfstartview=FitH, urlcolor=darkblue, linktocpage}

\newcommand{\arXhref}[1]{\href{http://arxiv.org/abs/#1}{#1}}

\begin{document}

\begin{flushright}
ZU-TH-8/16
\end{flushright}

\hfill

\vspace{1.0cm}

\begin{center}
{\LARGE\bf
Higgs mass and unified gauge coupling\\[5mm]
 in the NMSSM with Vector Matter
}
\\ \vspace*{0.5cm}

\bigskip\vspace{1cm}{
{\large \mbox{Riccardo Barbieri$^{a}$, Dario Buttazzo$^{b}$, Lawrence J. Hall$^{c}$, and David Marzocca$^{b}$} }
} \\[7mm]
  {\em $(a)$ Institute of Theoretical Studies, ETH Z\"urich, CH-8092   Z\"urich, Switzerland}\\
    {\em and  Scuola Normale Superiore, Piazza dei Cavalieri 7, 56126 Pisa, Italy}\\
  {\em $(b)$  Physik-Institut, Universit\"at Z\"urich, CH-8057 Z\"urich, Switzerland}  \\ 
  {\em $(c)$  Berkeley Center for Theoretical Physics, Department of Physics,} \\
    {\em and Theoretical Physics Group, Lawrence Berkeley National Laboratory,} \\
    {\em University of California, Berkeley, CA 94720, USA}

\end{center}
\vspace*{1.5cm}

\centerline{\large\bf Abstract}

We consider the  NMSSM extended to include one vector-like family of quarks and leptons.  If (some of) these Êvector-like matter particles, as  the Higgs doublets, have   Yukawa couplings Êto the singlet S that exceed unity at about the same scale $\Lambda \lesssim 10^3$ TeV, this gives the order $40\%$ enhancement of the tree level  Higgs boson mass required in the MSSM to reach 125 GeV. It is conceivable that the Yukawa couplings to the singlet S, although naively blowing up close to $\Lambda$, will not spoil gauge coupling unification. In such a case the unified coupling $\alpha_X$ could be interestingly led to a value not far from unity, thus providing a possible explanation for the number of generations. The characteristic signal is 
an enhanced resonant production of neutral spin zero particles at  LHC, that could even explain the putative diphoton resonance hinted by the recent LHC data at 750 GeV.

\vspace{0.3cm}

\begin{quote} \small

\end{quote}

\newpage

\section{Introduction}
\label{sec:intro}

The success of the Standard Model leads to several key questions; in this paper we are motivated by
\begin{itemize}
\item What determines the amount of matter; for example the number of generations $N_G$?
\item What determines the mass of the Higgs boson?
\end{itemize}

A possible answer to the first question is that the value of the unified gauge coupling at mass scales not far below the Planck scale is large, indicating the onset of semi-perturbative \cite{Babu:1996zv} and even non-perturbative behavior\cite{Maiani:1977cg, Cabibbo:1982hy, Maiani:1986cp}. 
In non-supersymmetric theories, with additional matter at the TeV scale, this implies matter contributions to the Standard Model beta functions equivalent to nine chiral generations, $N_G=9$.  This could be arranged into three chiral generations and three vector generations, but there are clearly many ways to arrange the new matter into complete $SU(5)$ multiplets.  In the Standard Model it is remarkable that the Higgs quartic coupling vanishes at a scale of order $10^{11}$ GeV; suggesting a possible line of attack to understand the size of the Higgs mass.  However, the addition of extra matter at the TeV scale changes the evolution of the quartic coupling removing the Higgs instability \cite{Dermisek:2012as}, so that some other understanding of the Higgs mass is needed.

In theories with TeV scale supersymmetry, it is striking that if $N_G < 5$ gauge coupling unification is highly perturbative, while with $N_G >5$ the gauge couplings become non-perturbative at or below $10^{10}$ GeV, destroying the success of perturbative gauge coupling unification.
The case of $N_G = 5$ is uniquely selected, and leads to unification at a scale of order $10^{17}$ GeV  and a unified coupling close to unity.  There are only two additions to the matter of the MSSM that yield such gauge running: one vector generation, $10 + \overline{10} + 5 + \overline{5}$, and  three/four copies of a vector fundamental $5 + \overline{5}$, with details dependent on threshold corrections and Yukawa couplings of vector matter.

In the MSSM the tree-level prediction for the Higgs boson mass is $M_Z |\cos 2 \beta|$.  As a zeroth order result this is highly successful, and motivates a continued emphasis on TeV scale supersymmetry.  Of course, a key question becomes the origin of the order 40\% enhancement required to reach 125 GeV.  One possibility is from loops of top squarks with a large mixing parameter; another is the addition of a superpotential interaction coupling the two Higgs doublets to a gauge singlet field $S$, $\lambda S H_u H_d$.  If the theory is perturbative to unified scales then RG scaling in the IR severely limits this contribution to the Higgs mass.  On the other hand if $S$ and/or $H_{u,d}$ are composites of some new interaction at scale $\Lambda$, as in Fat Higgs \cite{Harnik:2003rs, Chang:2004db} and $\lambda-$susy \cite{Barbieri:2006bg} schemes, the resulting contribution to the Higgs mass is typically (but not necessarily) too large, at least if $\Lambda < 10^8$ GeV.

\begin{figure}[t]
\begin{center}
\includegraphics[scale=0.70]{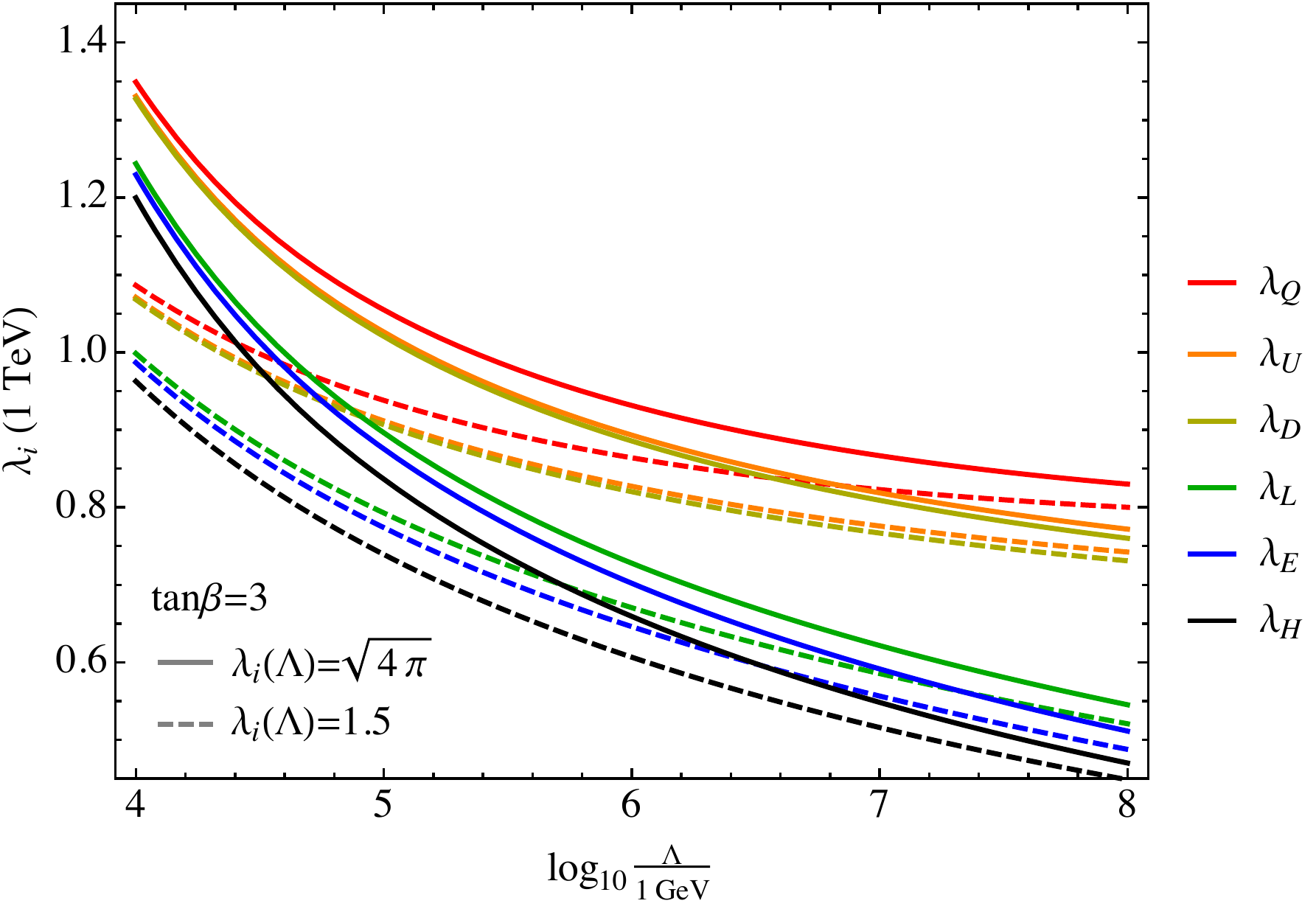}
\end{center}
\caption{Predictions for $\lambda_{H,i}$(TeV) when $\lambda_{H,i}(\Lambda)$ are large, for the case of one vector generation.}
\label{fig:lambdaTeV}
\end{figure}

In this paper we study supersymmetric theories that have gauge coupling unification with $N_G = 5$ and have $S$, Higgs and possible other states composite at scale $\Lambda \sim 10-10^3$ TeV.  We stress that a brief energy interval with strong dynamics at $\Lambda$ does not invalidate precision gauge coupling unification \cite{Hardy:2012ef}.   We are encouraged that in the Fat Higgs theory the constituents of the Higgs doublets have the same contribution to the running of the electroweak gauge couplings as the composites, $H_{u,d}$.  Some of the vector matter may acquire mass at scale $\Lambda$, for example to generate Yukawa couplings of the composite Higgs to quarks and leptons, but we assume that much of it does not.  

In the next section we define the theory below $\Lambda$ and discuss its relevant parameters.  In Section~\ref{sec:HiggsSector} we give results for symmetry breaking and the Higgs spectrum.  In Section~\ref{sec:Diphotons} we demonstrate that the 750 GeV diphoton signal seen  recently at the LHC can result from the production of the pseudoscalar $P$, and in Section~\ref{sec:VV} we argue that production of the CP-even component of $S$ leads to a $ZZ$ signal that can be significantly probed by running at $\sqrt{s} = 13-14$ TeV.  

\section{The Theory Below $\Lambda$}
\label{sec:theory}

We take the effective theory below $\Lambda$ to be described by the scale invariant superpotential\footnote{In a concise but self-evident notation, gauge symmetry and matter parity also allow the couplings $H\Phi\Phi, H\bar{\Phi}\bar{\Phi}, H\Psi\Phi, S\Psi\bar{\Phi}$, where $\Psi$ is the standard matter 15-plet. Only the last two couplings are strongly bound by flavour. Their smallness may be attributed to separate parities of the $\Psi$ and $\Phi$ fields.}
\be
W_{eff} \, =  \, W_{Yuk} + \lambda_H \, S H_u H_d + \lambda_i \, S \bar{\Phi}_i \Phi_i  +\frac{\kappa}{3} \, S^3~,
\label{eq:W}
\ee
where $W_{Yuk}$ are the Yukawa interactions of the Higgs doublets to quarks and leptons, and $(\Phi_i, \overline{\Phi}_i)$ are multiplets of vector matter. For the numerical work, in the following we choose these states to form one vector-like generation ($10 + \overline{10} + 5 + \overline{5}$), and in the Conclusions, Sec.~\ref{section:con}, we mention how this may be consistent with gauge coupling unification even if further vector matter resides near $\Lambda$.

\begin{figure}[t]
\begin{center}
\includegraphics[scale=0.70]{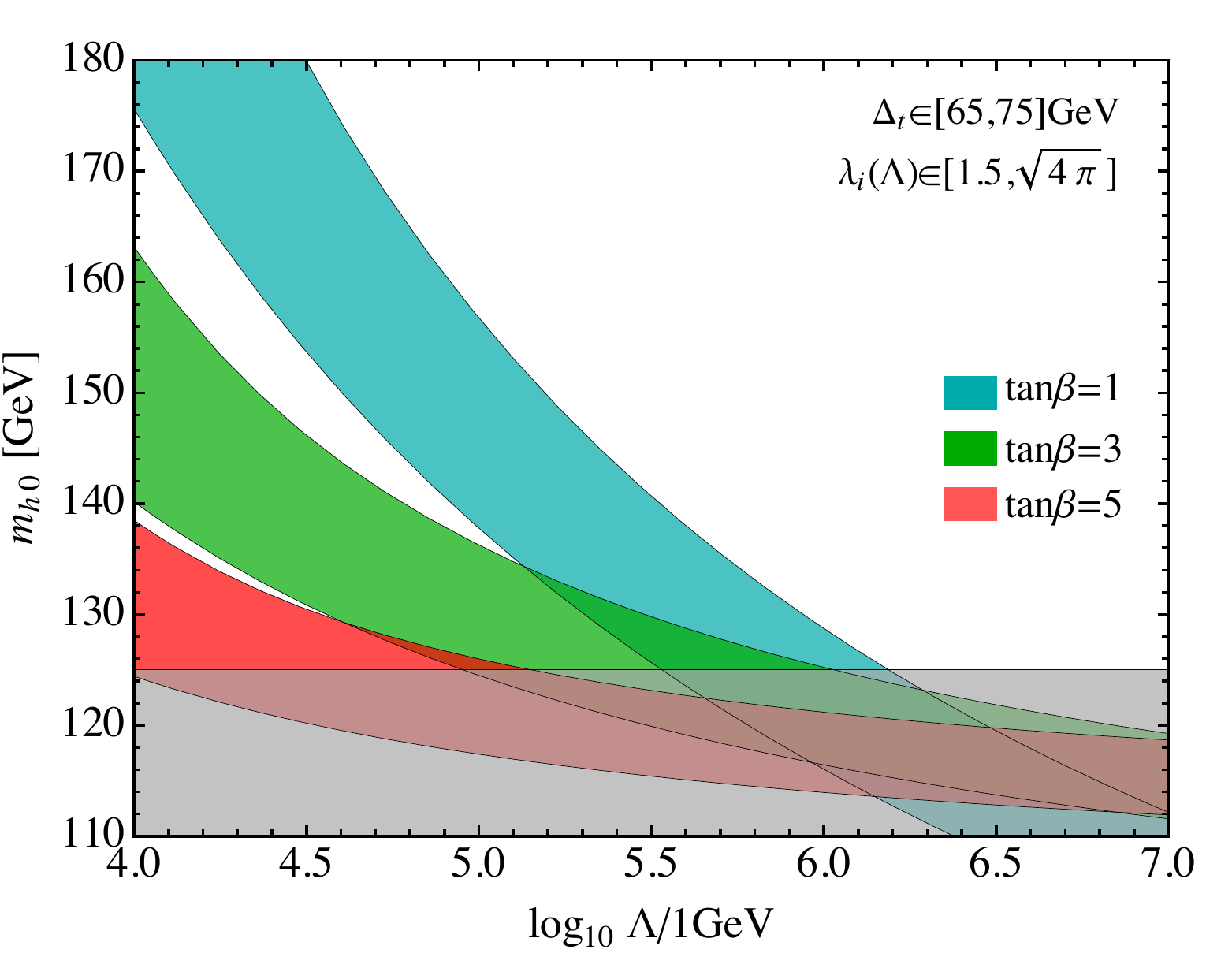}
\end{center}
\caption{The lightest Higgs mass before doublet-singlet mixing, which suppresses the mass, as a function of the scale of strong interactions, $\Lambda$, with $\lambda_{H,i}(\Lambda)$ large.}
\label{fig:Higgsmass}
\end{figure}

A key feature of this theory is the renormalization group flow of the couplings $(\lambda_H, \lambda_i, \kappa)$. With large boundary values at $\Lambda$, the couplings $\lambda_{H,i}(E)$ rapidly become insensitive to the boundary values and scale as $1/\ln (\Lambda/E)$ at $E \ll \Lambda$.  Fig.~\ref{fig:lambdaTeV} shows $\lambda_{H,i}$(TeV) as a function of $\Lambda$.  In particular the large number of fields coupling to $S$ insures that $\lambda_H$ and $\lambda_i$ drop faster in the IR than does $\lambda$ in the theory without  $(\Phi_i, \overline{\Phi}_i)$, so that the Higgs mass enhancement is typically of the required size, as illustrated by Fig.~\ref{fig:Higgsmass}.  The quantity $m_{h0}$ is the smallest, purely doublet, CP-even Higgs mass parameter, given by
\be
 m_{h0}^2 \, = \, M_Z^2 \cos^2 2\beta + \lambda_H^2 v^2 \sin^2 2 \beta + \Delta_{t}^2~,
 \label{eq:mh0}
 \ee
 where $ \Delta_t^2$ is the loop contribution coming mainly from virtual top quarks and squarks.  Each band in Fig.~\ref{fig:Higgsmass} comes in part from the range in $\lambda_i(\Lambda)$, the dominant effect at $\Lambda = 10\div 100$ TeV, and in part from the range in $\Delta_t = 65\div 75$ GeV,  corresponding to an average stop mass between 1 and 2.5 TeV, and a mixing mass below 1 TeV.
For small mixing mass and generally small $A$-terms, this range of $\Delta_t$ also includes radiative corrections proportional to powers of $\lambda_H(1 \TeV)$ (see App.~C of Ref.~\cite{Ellwanger:2009dp}).
As we show in detail shortly, mixing with the singlet scalar leads to a physical Higgs mass somewhat less than $m_{h0}$.

In the MSSM, ignoring CP violation, there are four parameters that enter the electroweak symmetry breaking sector, $(\mu; m_u^2, m_d^2, B)$; after minimization the four independent parameters can be taken to be $(\mu; v, \tan \beta, m_A)$.  However, in addition three further parameters associated with the top squark are needed to compute the Higgs mass.  Assuming CP conservation, the theory of (\ref{eq:W}), supplemented with the soft SUSY breaking terms defined in the next Section, has seven parameters associated with Higgs and $S$ vevs, $(\lambda_H, \kappa; m_u^2, m_d^2, m_s^2, A_\lambda, A_\kappa)$, and these translate into $(\lambda_H, \kappa; v, \tan \beta, v_s, m_A, m_P)$, with $\lambda_H$ depending only logarithmically on $\Lambda$.  For small top squark mixing, the dominant contributions to the Higgs mass arise from these parameters; we explore the resulting prediction in detail, finding that $\Lambda$ must be less than $10^3$ TeV, as is evident also from Fig.~\ref{fig:Higgsmass}.  The mass of the CP-even scalar in $S$, $m_S$, is not an independent parameter, but is given in terms of $(v_s, \kappa, m_P)$.

The masses of the Higgsinos and vector matter are given by $\mu = \lambda_H v_s$ and $M_i = \lambda_i v_s$, with ratios that are predicted and depend only on the gauge quantum numbers of $\Phi_i$.  In this paper we take the supersymmetry breaking scale and hence $v_s$ to be order TeV. These masses are therefore proportional to the couplings $\lambda_{H,i}(1 \TeV)$ shown in Fig.~\ref{fig:lambdaTeV}.

\begin{figure}[t]
\begin{center}
\includegraphics[scale=0.50]{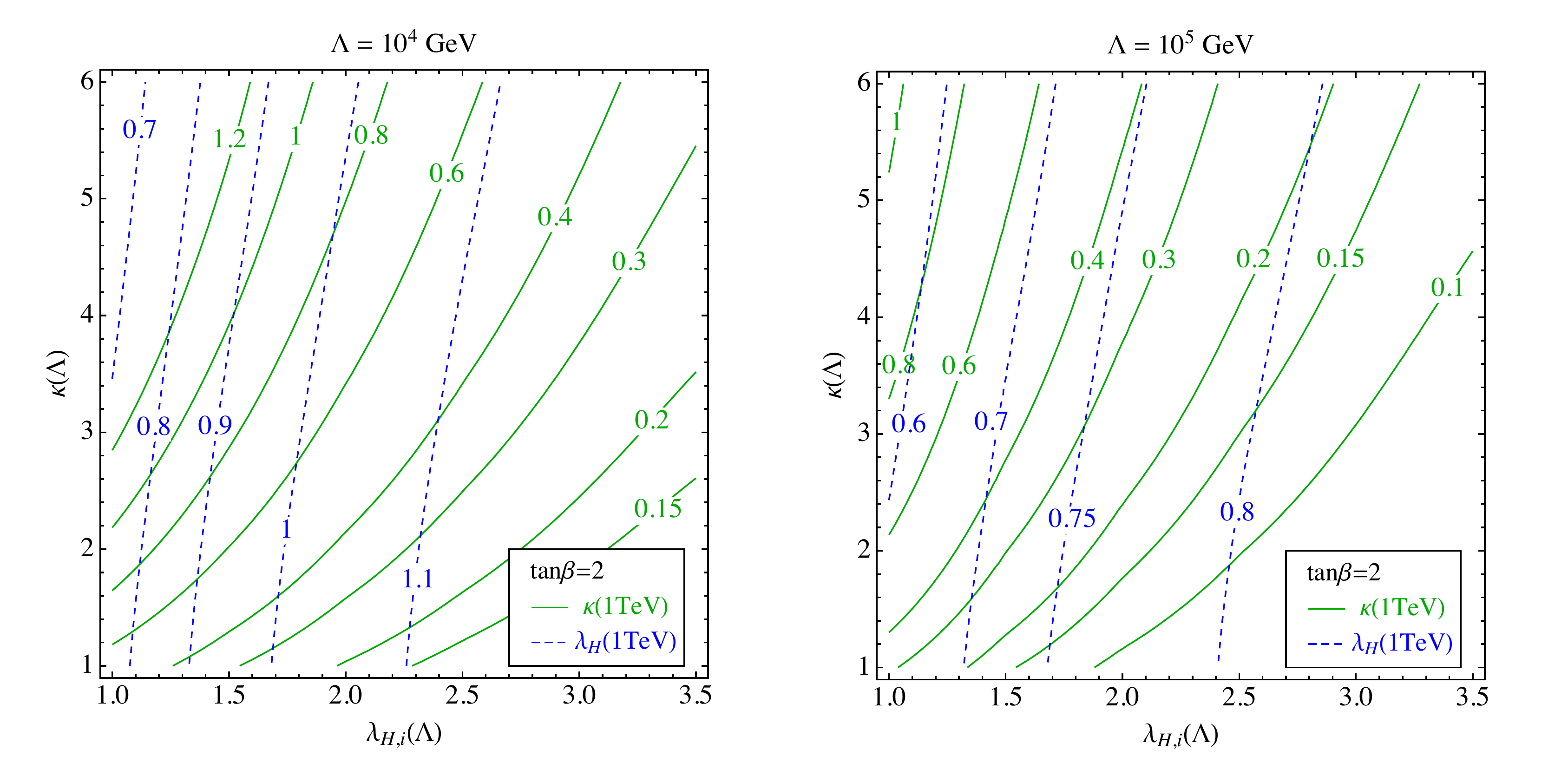}
\end{center}
\caption{Low energy values of  $\kappa$ and $\lambda_H$ as functions of their initial conditions at $\Lambda= 10^4 \GeV$ (left) and $ \Lambda= 10^5$ GeV (right). All the $\lambda_i$ have the same initial condition as $\lambda_H$. }
\label{fig:Boundary_conditions}
\end{figure}

As shown in Fig.~\ref{fig:Higgsmass}, the Higgs mass $m_{h0}$ before doublet-singlet mixing  depends predominantly, other than $\tan{\beta}$, on $\Lambda$, since $\lambda_H(E)$ becomes  rapidly insensitive to the boundary value. The physical Higgs mass, as the entire scalar spectrum, however, crucially depends as well on the other dimensionless parameter appearing in the superpotential term $(\kappa/3) S^3$. Furthermore this same term is what prevents the vev $v_s$ of the field $S$ from running to infinity. As such the low energy value of $\kappa$ cannot be too small. 

Fig.~\ref{fig:Boundary_conditions} shows the values at 1 TeV of $\kappa$ and $\lambda_H$ as functions of their initial values at $\Lambda= 10^4, 10^5$ GeV. Whereas the variation of  $\lambda_H(1\TeV)$ with the initial conditions is relatively weak even at low values of $\Lambda$, this is not the case for $\kappa (1\TeV)$, due to the cubic dependence on $S$ of the corresponding superpotential term. The renormalization of the $S$ field leads to $\kappa$ decreasing in the IR more rapidly than $\lambda_H$ so that at the TeV scale it is typically smaller.  As such $\kappa (1\TeV)$  takes a value that is sensitive to the boundary values and becomes a relevant parameter in the entire Higgs potential and Higgs spectrum.  From now on $\lambda_H(1\TeV)$ and $\kappa (1\TeV)$ are denoted by $\lambda_H$ and $\kappa$ unless differently stated.

\section{Symmetry Breaking and the Higgs Spectrum }
\label{sec:HiggsSector}
The Higgs potential and Higgs spectrum are determined by the $\lambda_H$ and the $\kappa$ terms in the superpotential, eq.~\eqref{eq:W}, together with the soft SUSY-breaking potential, dependent on the corresponding scalar fields, for which we use the same notation
\begin{equation}
V_{soft}= m_s^2 |S|^2 + m_u^2 |H_u|^2 + m_d^2 |H_d|^2 + ( A_\lambda \lambda_H S H_u H_d + A_\kappa 
\frac{\kappa}{3} S^3 + h.c.).
\end{equation}
This potential is extensively studied in the literature\cite{Ellis:1988er, Ellwanger:2009dp}. In a range of the parameters  
\begin{equation}
(\lambda_H, \kappa; m_u^2, m_d^2, m_s^2, A_\lambda, A_\kappa),
\end{equation}
 it has a CP-conserving 
$SU(2)\times U(1)$-breaking minimum. For our purposes the physical Higgs spectrum and the relevant mixing angles are  more effectively described in terms of a different choice of parameters
$(\lambda_H, \kappa; v, \tan \beta, v_s)$ and two physical masses themselves: $m_A, m_P$, with $A$ and $P$ the two neutral CP-odd scalars.

By expanding in $v/v_s$ it is straightforward to obtain all the other scalar masses as well as the composition of all the Higgs states. First the neutral CP-odd states themselves, $A$ and $P$. In terms of the real and imaginary parts of the various fields, $S= v_s + (S_R + i S_I)/\sqrt{2}$, $H^0_u = v_u + (H_{uR}+i H_{uI})/\sqrt{2}$, $H^0_d = v_d + (H_{dR}+i H_{dI})/\sqrt{2}$, it is
\begin{equation}
P= \cos\theta_P S_I + \sin\theta_P (c_\beta H_{uI} + s_\beta H_{dI})\quad\quad
A = -\sin\theta_P S_I +  \cos\theta_P (c_\beta H_{uI} + s_\beta H_{dI})~,
\end{equation}
where $s_\beta = \sin \beta$, $c_\beta = \cos \beta$, and, to leading order in $v/v_s$,
\begin{equation}
\sin^2\theta_P = \frac{\lambda_H^2 v^2 \mu^2}{(m_A^2-m_P^2)^2} \left[ s_\beta c_\beta \left(\frac{m_A}{\mu}\right)^2 -3 \frac{\kappa}{\lambda_H}\right]^2, \quad\quad \mu = \lambda_H v_s~.
\label{eq:Pmixing}
\end{equation}
We have taken $P$ as predominantly singlet under $SU(2)\times U(1)$.

\begin{figure}[t]
\begin{center}
\includegraphics[scale=0.45]{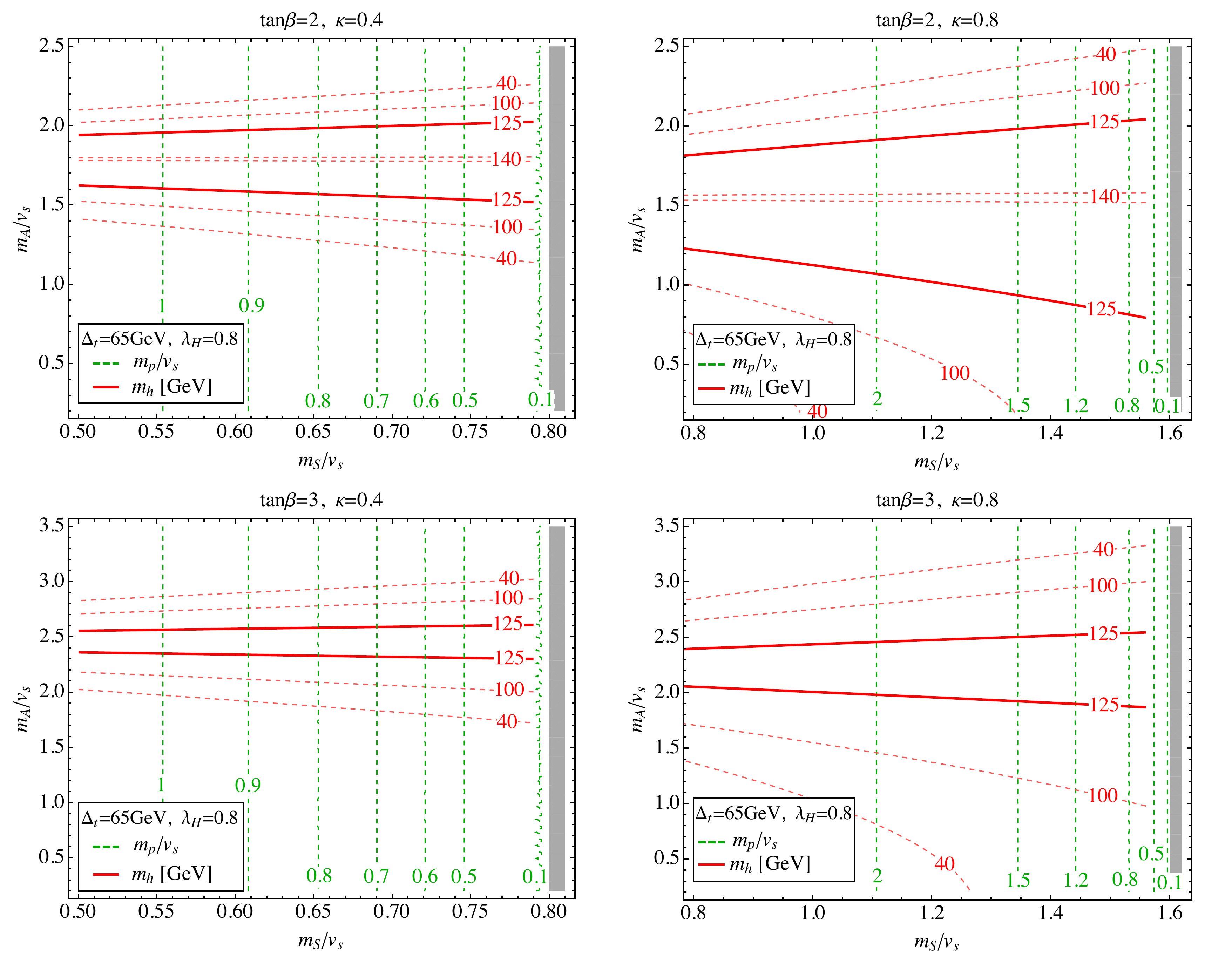}
\end{center}
\caption{Isolines of constant $m_h$ (solid and dashed red) and $m_P/v_s$ (dashed green) after inclusion of the doublet-singlet mixing, in the plane $(m_S/v_s, m_A/v_s)$.}
\label{fig:Higgs_mass_kappa0408}
\end{figure}

Denoting with $(h, H, S)$ the neutral CP-even states\footnote{With an abuse of notation for the state $S$, not to be confused with the complex field $S$ used so far}, their masses  and composition are, to a sufficient level of approximation,
\begin{equation}
m_h^2 = m_{h0}^2 - \Delta m_h^2,\quad m_H^2 = m_A^2,\quad m_S^2 = \frac{1}{3}\left(12\kappa^2 v_s^2- m_P^2\right) ,
\end{equation}
\begin{eqnarray}
h &=& \cos\theta_S (c_\beta H_{dR} + s_\beta H_{uR}) + \sin\theta_S S_R, \nonumber \\
S &=&  -\sin\theta_S (c_\beta H_{dR} + s_\beta H_{uR}) + \cos\theta_S  S_R , \\
H &=& -s_\beta H_{dR} + c_\beta H_{uR} , \nonumber
\end{eqnarray}
where
\begin{equation}
\Delta m_h^2 =  4 \lambda_H^2 v^2 \Big(\frac{\mu}{m_S}\Big)^2 \left[ 1 - s_\beta c_\beta \left(s_\beta c_\beta \left(\frac{m_A}{\mu}\right)^2 + \frac{\kappa}{\lambda_H}\right)\right]^2
\label{eq:HiggsMassMix}
\end{equation}
and
\begin{equation}
\sin^2{\theta_S} = \frac{\Delta m_h^2}{m_S^2}.
\label{thetaS}
\end{equation}
In the same approximation, the charged Higgs state $H^\pm$ is degenerate with $H$ and $A$.

As anticipated, the doublet-singlet mixing described by the angle $\theta_S$ corrects downward by the amount $\Delta m_h^2$ the mass $m_{h0}^2$ of the physical Higgs boson, shown in Fig.~\ref{fig:Higgsmass}. Fig.~\ref {fig:Higgs_mass_kappa0408}  illustrates the range of parameters needed to obtain the observed value $m_h = 125$ GeV. The mixing correction to $m^2_{h0}$ vanishes when the argument in the parenthesis in Eq.~\eqref{eq:HiggsMassMix} is zero. This tuned region corresponds to the maximal Higgs mass in each panel in Fig.~\ref {fig:Higgs_mass_kappa0408}. For other values of the parameters, in particular larger or smaller $m_A/v_s$, the correction to the Higgs mass becomes sizable and negative. Note that neither case $m_A/v_s \ll 1$ nor $m_A/v_s \gg 1$ corresponds to a decoupling regime.
As further shown in the following Sections, we consider the general consistency of this range of parameters, with $v_s$ close to 1 TeV, as positive evidence for the model under consideration.

\section{Putative 750 GeV Diphoton Signal }
\label{sec:Diphotons}

\begin{figure}[t]
\begin{center}
\includegraphics[scale=0.55]{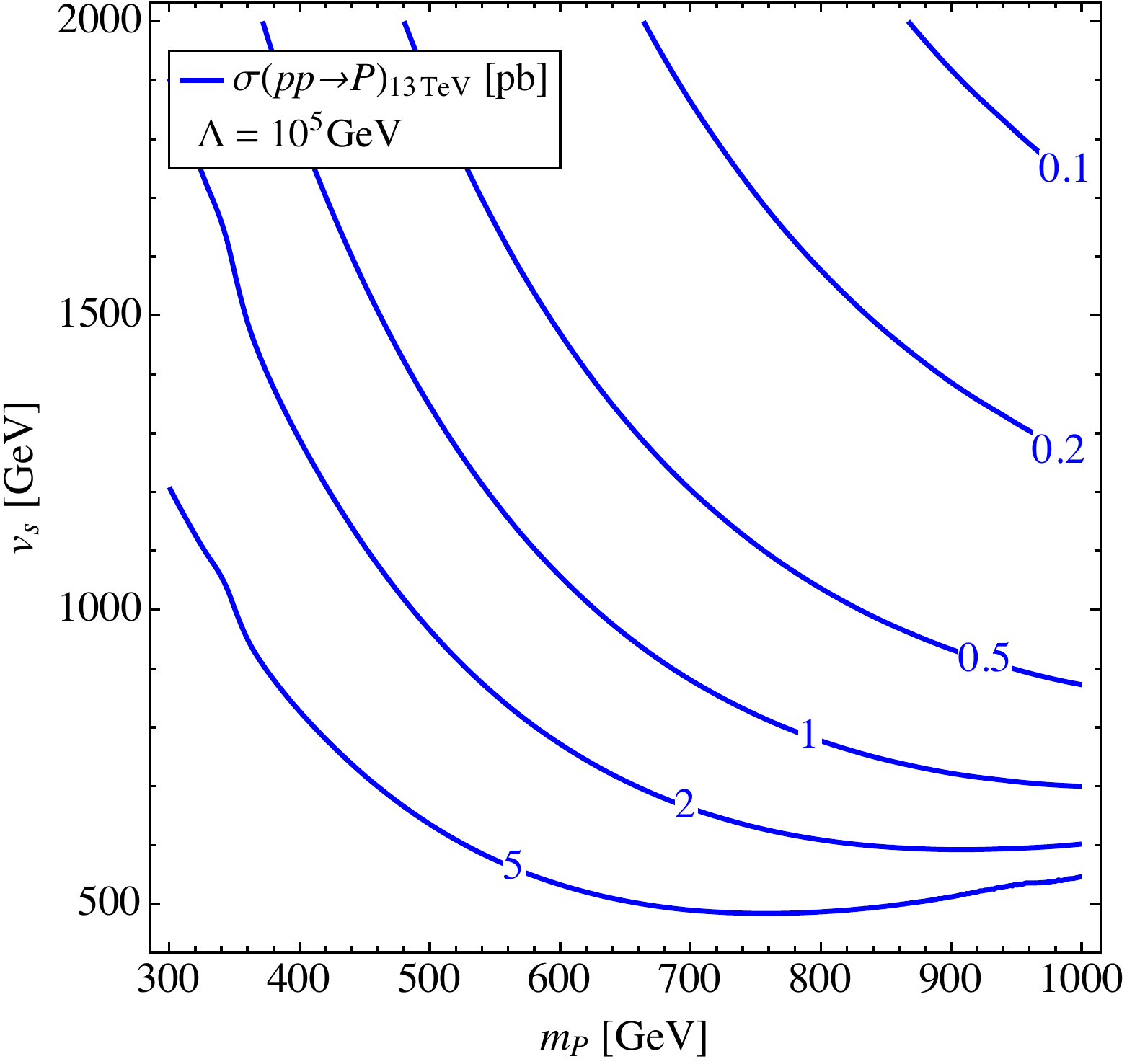}
\end{center}
\caption{Gluon-fusion production cross section in pb of the $P$ state at 13 TeV, as function of its mass, $m_P$, and of the S-vev, $v_s$. For definiteness we fix $\Lambda = 10^5 \GeV$ and $\lambda_i(\Lambda)=1.5$.}
\label{fig:P_prodXS_13TeV}
\end{figure}

At least for the states $S$ and $P$, which are predominantly singlets under $SU(2)\times U(1)$, a clear consequence of the picture described so far is an enhanced resonant production at the LHC through the gluon-fusion channel, as shown in Fig.~ \ref{fig:P_prodXS_13TeV} for the CP-odd state $P$, obtained by rescaling the NNLO QCD cross section of a SM-like Higgs from Ref.~\cite{Heinemeyer:2013tqa}. This suggests to take the putative resonance hinted by the recent LHC data at 750 GeV \cite{ATLAStalk,CMStalk} as an illustrative case.

This result received an overwhelming response from the theoretical community, and a large number of preprints appeared on the subject\footnote{To our knowledge, the model studied in Ref.~\cite{Tang:2015eko}, that considers the possibility of interpreting the  putative resonance at 750 GeV  in the NMSSM with vector-matter as a $P$-like state decaying into two photons, is the closest to our analysis. Different NMSSM scenarios were studied at least in \cite{Ellwanger:2016qax, Domingo:2016unq}. In the context of the MSSM with an extra singlet, Refs.~\cite{Hall:2015xds,Tang:2015eko,Dutta:2016jqn,Han:2016fli} consider extra vector-like matter in $SU(5)$ multiplets while a single vector-like quark is added in Ref.~\cite{Wang:2015omi}. Ref.~\cite{King:2016wep} studies the excess in the context of $E_6$ unification, and Ref.~\cite{Gabrielli:2015dhk} aims at describing the signal without extra matter.}. In particular, among the first works which provided a combination of the experimental results, as well as interpretations in a selection of simple models relevant to our case, were Refs.~\cite{Buttazzo:2015txu,Franceschini:2015kwy,DiChiara:2015vdm,Ellis:2015oso,Gupta:2015zzs,Falkowski:2015swt,Low:2015qep,Bellazzini:2015nxw,Becirevic:2015fmu}. The fits in these works show that assuming production via gluon fusion and  combining LHC data from Run-1 and Run-2 gives a rate for the diphoton signal at 13 TeV between approximately $3$ and $10$ fb. For definiteness, in the following we use the result of Ref.~\cite{Buttazzo:2015txu}:
\be
	\mu_{13 \TeV}(\gamma\gamma) = (4.6\pm 1.2)~ \text{fb}~.
	\label{eq:diphoton_rate}
\ee

In the present context the gluon-gluon production cross section for the $S$ state is similar to the one for $P$, only rescaled down by about a factor of $\frac{9}{4}$, neglecting scalar contributions in the loop. What differs between $S$ and $P$ are their Branching Ratios, since the opposite CP-nature allows only in the $S$ case a tree level coupling to the $WW, ZZ, hh$ pairs. The CP-odd state is therefore favorite to have a sizable Branching Ratio in $\gamma\gamma$, with $t\bar{t}$ only as competing channel.

\begin{figure}[t]
\begin{center}
\includegraphics[scale=0.55]{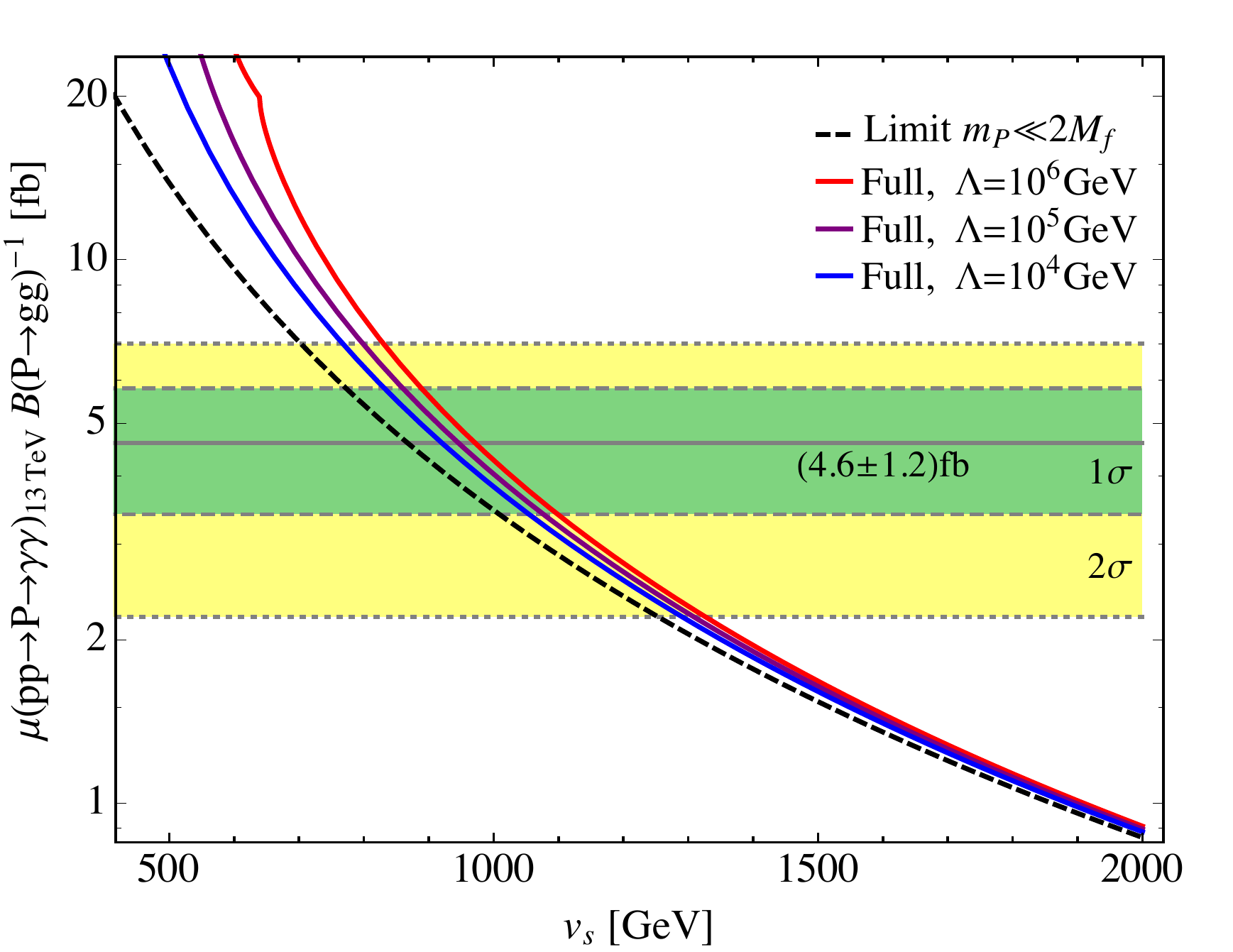}
\end{center}
\caption{Signal rate $\mu{(pp\rightarrow P\rightarrow \gamma\gamma)}$ at 13 TeV as a function of $v_s$, normalized to the $\BR(P\rightarrow gg)$, for different values of $\Lambda$ (with $\lambda_i(\Lambda)=1.5$). Also shown are the $1\sigma$ and $2 \sigma$ bands of the estimated diphoton signal.}
\label{fig:rate_vevS}
\end{figure}

\begin{figure}[t]
\begin{center}
\includegraphics[scale=0.45]{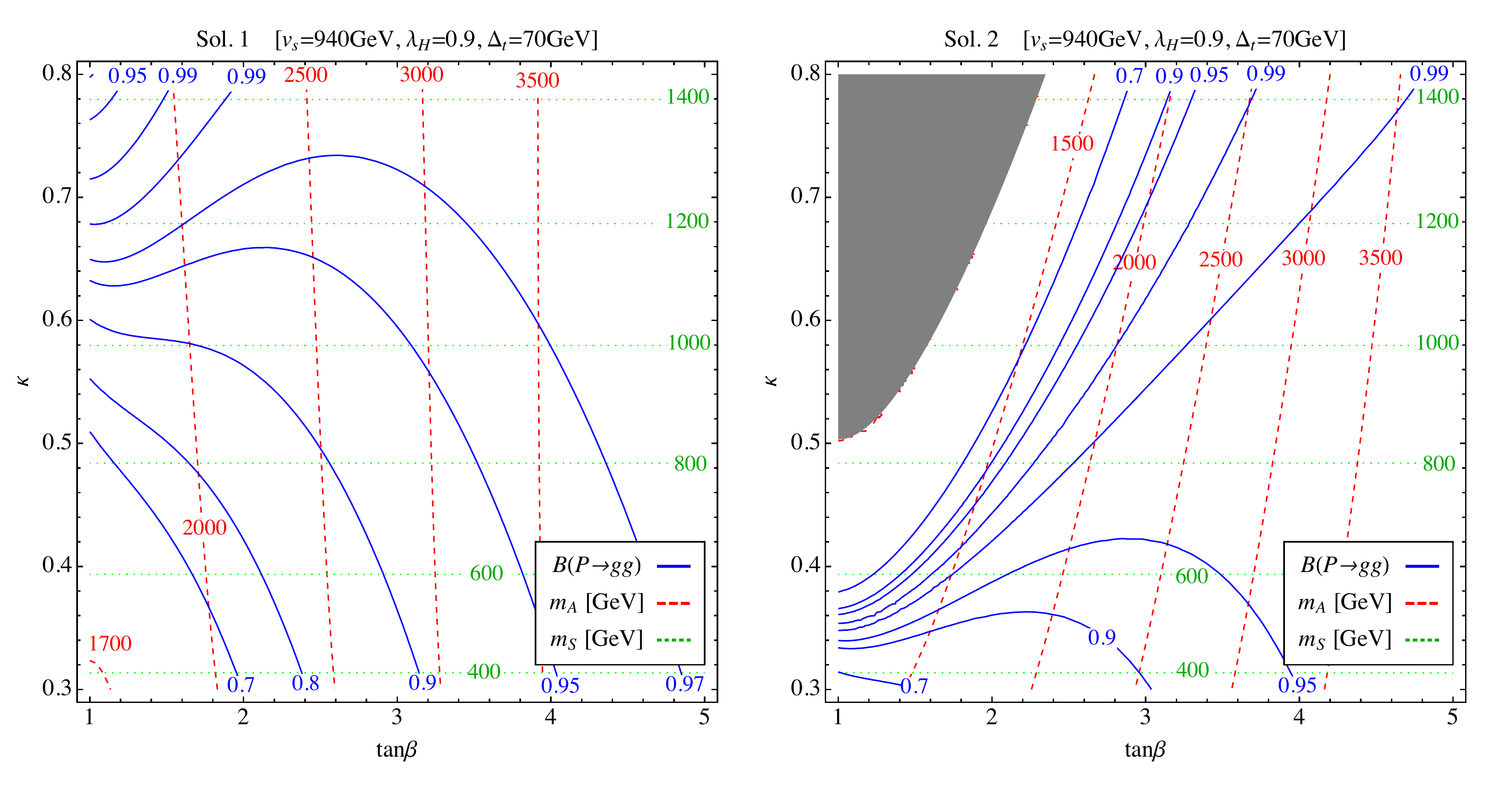}
\end{center}
\caption{Isolines of $\BR(P\rightarrow gg)$ (solid blue) for the two possible values of $m_A$ in Fig.~\ref{fig:Higgs_mass_kappa0408}.  (left, higher $m_A$; right, lower $m_A$). Also shown are isolines of $m_S$ (dotted green) and $m_A$ (dashed red).}
\label{fig:tanbeta_k_vS940}
\end{figure}

Taking $m_P=750$ GeV, the width $\Gamma(P\rightarrow \gamma \gamma)$ depends in principle on the $\lambda_{H,i}$ and on the masses $M_f = (\mu, M_i) = \lambda_{H,i}v_s$, hence on $v_s$ and $\Lambda$. The dependence on $\Lambda$, however, is relatively weak, since for $2 M_f\gg m_P$ the width goes as $\lambda_{H,i}^2/M_f^2$, and the dependence on the couplings drops away. In Fig.~\ref{fig:rate_vevS} we show the signal $\mu(pp\rightarrow P\rightarrow \gamma\gamma)$ at $13$ TeV, divided by $\BR(P\rightarrow gg)$, a combination which only depends on $\Gamma(P\rightarrow \gamma \gamma)$. In turn we show in Fig.~\ref{fig:tanbeta_k_vS940} $\BR(P\rightarrow gg)$ itself, which can deviate from one only due to the competition from
\begin{equation}
\Gamma(P\rightarrow t\bar{t})= \sin^2\theta_P \frac{3 G_F m_t^2}{4\sqrt{2}\pi \tan^2\beta} m_P~.
\label{eq:Pttbarwidth}
\end{equation}

This width, hence $\BR(P \to gg)$, is controlled by $\theta_P$, eq.~\eqref{eq:Pmixing}, which depends on a set of parameters constrained by $m_h = 125$ GeV, as shown in Fig.~\ref{fig:Higgs_mass_kappa0408}. In these figures, the two possible values of $m_A$ which reproduce the correct Higgs mass correspond to the two solutions represented in Fig.~\ref{fig:tanbeta_k_vS940}. The dark region in Fig.~\ref{fig:tanbeta_k_vS940} (right) corresponds to $\sin\theta_P > 0.3$, that goes beyond the approximation for small $v/v_s$.

As evident from this figure, the decay $P\rightarrow t\bar{t}$ plays a limited role and only in a small corner of the parameter space. Therefore, Fig.~\ref{fig:rate_vevS} shows that the $P$ state could account for the signal at 750 GeV with the estimated rate in eq.~\eqref{eq:diphoton_rate} for $v_s \sim 0.8 \div 1.2 \TeV$. 
Other interesting decay rates are
\begin{equation}
\frac{\Gamma (P\rightarrow WW, ZZ, Z\gamma)}{\Gamma (P\rightarrow \gamma\gamma)}=
(5.3\div 6.3, 2.0\div 2.4, 0.26\div 0.31),
\end{equation}
i.e, normalizing to a signal rate $\mu_{13 \TeV}(\gamma\gamma) = 4.6$ fb, 
\begin{equation}
\mu_{13 \TeV}(WW, ZZ, Z\gamma) = (24\div 29, 9.2\div 11, 1.2\div 1.5) \text{fb}.
\end{equation}
The total width of the $P$ state never exceeds 100 MeV. This prediction of our model is in some tension with the ATLAS 13\,TeV diphoton analysis \cite{ATLAStalk}, which shows a mild preference for a large-width $\Gamma \sim 45 \GeV$. Even though it is still too early to draw definite conclusions on such properties of the excess, our model could be refuted if the resonance at 750 GeV  were confirmed with a large width in future analyses.

\section{The $WW$ and $ZZ$ Signal}
\label{sec:VV}

As anticipated, the CP-even state $S$ is broader due to its decays into a pair of $WW, ZZ, hh$, with approximate relative rates $2\div 1\div 1$, since
\begin{equation}
\Gamma(S\rightarrow ZZ)=   \sin^2\theta_S \frac{ G_F m_S^3}{16\sqrt{2}\pi}~.
\end{equation}
This may clearly give rise to another interesting signal. One may actually wonder if data collected at 8 TeV do not already represent a relevant constraint in the parameter space. Note that the $\BR(S\rightarrow t\bar{t})$ never exceeds the $10\div 20\%$ level, so that $\BR(S\rightarrow ZZ) \approx 1/4$.\footnote{We are not including the effect of the mixing of $S$ with $H$, which could somewhat increase $\Gamma(S\rightarrow t\bar{t})$.} The total width of the $S$ state ranges from about $0.5$ GeV up to about $10$ GeV, mostly depending on the value of $\kappa$.

\begin{figure}[t]
\begin{center}
\includegraphics[scale=1]{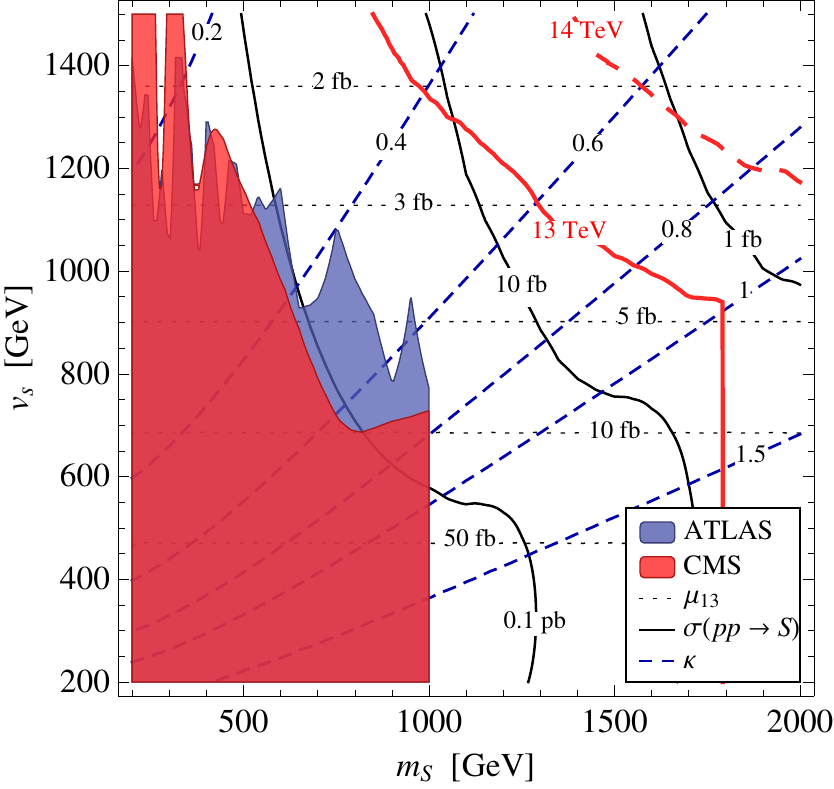}
\end{center}
\caption{Isolines of $\sigma(pp\rightarrow S)$ at 8 TeV (black lines). The red and blue regions give the exclusion from $S \to ZZ$ by current data at 8 TeV. The red solid (dashed) line gives an estimate of the future sensitivity on this channel at 13 (14) TeV with $100~ (300)~\text{fb}^{-1}$ of integrated luminosity, taken from Ref.~\cite{Buttazzo:2015bka}. Also shown is $\mu_{13\TeV}(P \to \gamma\gamma)$ normalized to $\BR(P\to gg)$ (dotted black lines) and the parameter $\kappa$ (dashed blue lines).}
\label{fig:ZZ_exclusion_k}
\end{figure}

\begin{table}[b]
\begin{center}\vspace{0.5cm}
\begin{tabular}{c|c|c|c|c|c} \hline
$\Lambda$ & $\tan\beta$ & $v_s$ & $\kappa$ & $m_S$ & $m_A$ \\ \hline
$10^{4\div 5} \GeV$ & $1\div 3$ & $0.8\div 1.2 \TeV$ & $0.4\div 1$ & $0.8\div2 \TeV$ & $1\div 3 \TeV$
\end{tabular}
\caption{\label{tab:region} Parameter range of the model, favoured by the Higgs mass, the diphoton excess, and the bound from $S\to ZZ$.}
\end{center}
\end{table}

The relevant $8 \TeV$ analyses of the $S\to ZZ$ channel are \cite{Aad:2015kna} (ATLAS) and \cite{Khachatryan:2015cwa} (CMS). The current and projected exclusion from the $ZZ$ signal is shown in Fig.~\ref{fig:ZZ_exclusion_k}. The combination of Figs.~\ref{fig:rate_vevS}, \ref{fig:tanbeta_k_vS940}, \ref{fig:ZZ_exclusion_k} and an estimated rate $\mu_{13\TeV}(\gamma\gamma) = (4.6\pm 1.2)~\text{fb}$ show that the 750 GeV signal could be reproduced in the parameter range summarized in Table~\ref{tab:region}.
The necessary low-energy value of $\kappa(1 \TeV) \simeq 0.4 \div 0.7$ requires some non-generic boundary conditions $\kappa(\Lambda) \gtrsim \lambda_{H,i}(\Lambda)$, in particular for larger values of the scale $\Lambda$, as can be seen in Fig.~\ref{fig:Boundary_conditions}.

\section{Conclusions and Outlook}
\label{section:con}

 TeV-scale supersymmetry gives a leading contribution to the Higgs mass of about $M_Z$.  In this paper we have demonstrated that the required enhancement to 125 GeV, arising from the $SH_uH_d$ coupling and from $S/H$ mixing, is typical in the NMSSM with a scale-invariant superpotential and a generation of vector matter, as shown in Figs. \ref{fig:Higgsmass} and \ref{fig:Higgs_mass_kappa0408}.  The couplings of vector matter and Higgs to the singlet field become large at a scale $\Lambda \lsim 10^3$ TeV, leading to a highly predictive spectrum of Higgs and vector matter states at the scale of the singlet vev, taken to be $\simeq 1 \TeV$. 

Loops of vector matter lead to an enhanced production rate of the pseudoscalar $P$ and the scalar $S$ that are dominantly electroweak singlets, so that signals are expected at the LHC.  Fig.~\ref{fig:rate_vevS} shows that the observed 750 GeV diphoton excess
arises from $P$ production if the singlet vev is $\simeq 1 \TeV$, and Fig.~\ref{fig:tanbeta_k_vS940} shows that a significant signal for $P \rightarrow \bar{t}t$ may be present in part of parameter space.   Most important is that $S/H$ mixing, required for the Higgs mass, leads to a large signal for $S \rightarrow ZZ,WW$.  As shown in Fig.~\ref{fig:ZZ_exclusion_k}, bounds from Run 1 exclude some parameter space for $m_S < 1$ TeV, and data at $\sqrt{s}$ = 13, 14 TeV will provide a powerful probe of the theory.

The masses of vector matter, with ratios only dependent on the gauge quantum numbers (at least for moderate $H\Phi\Phi, H\bar{\Phi}\bar{\Phi}$ couplings), lie in the TeV range. With $m_P=750$ GeV quarks are between 0.8 and 1.5 TeV and leptons between 0.6 and 1.2 TeV. The heavier states cascade to the lighter ones, which ultimately decay promptly to standard quarks and leptons with the emission of $W, Z$, and $h$.

It is possible that the lightest R-parity odd particle, $\chi$, be an admixture of higgsino and singlino with a mass around one TeV. In this case, taking into account the effect of S-exchange in the determination of its relic abundance, $\chi$ is a candidate for Dark Matter in a suitable range of the parameter space characterized by 
$(\lambda_{H,i}, \kappa, \tan\beta, v_s)$.

The results summarized above are insensitive to the amount of vector matter below $\Lambda$, as long as it is substantial.   Our numerical results are for the case of a vector generation at the TeV scale, but adding further multiplets yields the same phenomenology with a rescaled $\Lambda$.  The case of perturbative gauge coupling unification with a single vector generation is particularly interesting, since only in this case is the amount of matter determined by an order unity unified gauge coupling.  However, although non-perturbative physics at $\Lambda$ yielding composite states can be made consistent with gauge coupling unification, some vector matter is required near $\Lambda$, leaving less than a generation below $\Lambda$.  We note that if the strong dynamics were conformal above $\Lambda$, with SU(5) as a global symmetry, precise gauge coupling unification might occur where the unified coupling becomes order unity with more than 1 generation of vector matter.  All of this requires further study to be made concrete.

\section*{Acknowledgments}
L. Hall thanks Keisuke Harigaya and Yasunori Nomura for many useful conversations. R. Barbieri, D. Buttazzo, and D. Marzocca thank Gino Isidori for interesting discussions on various points of this paper. This work was supported in part by the Director, Office of Science, Office of High Energy and Nuclear Physics, of the US Department of Energy under Contract DE-AC02-05CH11231 and by the National Science Foundation under grants PHY-1002399 and PHY-1316783. R. Barbieri wants to thank Dr. Max R\"ossler, the Walter Haefner Foundation and the ETH Zurich Foundation for support. D. Buttazzo and D. Marzocca are supported in part by the Swiss National Science Foundation (SNF) under contract 200021-159720.




\end{document}